\documentclass[aps,preprint,showpacs]{revtex4}
\begin{document}        %
\draft
\title{Note on "Hydrodynamic Phase Locking of \\ Swimming Microorganisms"}
\author{Guanghua Zhu} 
\affiliation{QingHai-Lake Centre, 2 Street Hua-Yuan South Xining
810000, China} %
%
\begin{abstract}
We make remarks on Elfring and  Lauga's [{\it Phys. Rev. Lett.}
{\bf 103}, 088101  (2009)] paper. The energy dissipation or
viscous dissipation  plays an important role in the phase-locked
state.
%
\end{abstract}
\maketitle
%
\bibliographystyle{plain}
Recently  "What is still not understood is what constitutes the
essential physical ingredients to obtain an evolution in time to a
phase-locked configuration between cells swimming close to each
other." was raised by Elfring and  Lauga [1]. Therein, Elfring and
Lauga considered a simplified model of nearby swimming cells with
a prescribed waveform. They showed that stable phase locking can
be obtained purely passively, due to hydrodynamic interactions.
They also claimed that "The phase-locked state to which the cells
evolve is dictated {\it solely by the geometry of the flagellar
waveforms of the cells} (specifically, their front-back
asymmetry), and  {\it not by considerations of energy
dissipation.}"\newline In fact, by following previous approach
[2], Elfring and  Lauga considered a model of coswimming cells
consisting of two infinite parallel 2D sheets propagating lateral
waves of transverse oscillations with prescribed wave number $k$,
frequency $\omega$, and wave speed $c=\omega/k$, each sheet is
thereby propelled in the direction opposite to the wave. The shape
of the waveform is presumed to be the same for both swimmers, and
is described by an arbitrary function $a$ [1].\newline As the
present author checked, there are some doubtful expressions in [1]
: (1) The boundary conditions for Eq. (1) should be changed to
$u(y_1)=1$, $u(y_2)=1-U_{\Delta}$ considering almost the same
mathematical treatment in [3] (Eq. (3) therein) together with Fig.
1 of [1]; (2) Eq. (1) in [1] should be read as
$u=[(y-y_1)(y-y_2)dp/dx]/2-U_{\Delta} (y-y_1)/(y_2-y_1)+1$
following the same reasoning (cf. Eq. (3) in [3]); (3) Eq. (3) in
[1] should be read as $dp/dx=12 (-I_2/(I_3 h^3)+1/h^2)$ with
$Q=I_2/I_3$ where $I_j$ follows the same definition in [1]. (4)
Other subsequent derivations related to Eq. (1) in [1] should be
carefully examined with the same thinking. \newline Meanwhile as
Elfring and  Lauga adopted two infinite, parallel, two-dimensional
waving sheets which might be flexible and undergo large amplitude
of motions, thus $y_2 - y_1 >0$ or $a(x+\phi)-a(x)+1>0$ should be
always satisfied! Otherwise  another kind of wave breaking could
occur [4]! In fact this checking also appears at the denominator
of the (right-hand-side) integrand of $\Lambda^{-1}$ of Eq. (12)
in [1]! $a(x+\phi)-a(x)+1=0$   is not allowed! Note that Eq. (12)
is related to the solving Eq. (11) in [1] and the presentation for
Fig. 3 of [1].
\newline The last remark is about the role of energy dissipation
or viscous dissipation which is underestimated in [1]. Considering
the dimensionless definition of the fluid force ($f$) and energy
dissipation rate per unit width ($\dot{E}$) [1], where the
viscosity ($\mu$) occurs ($\mu$ also appears in the dimensionless
definition of the pressure ($p$)), we have, near the in-phase
configuration ($\phi_0=0$), the force $|f_{x_{\theta}}|\sim \phi'
\delta^4$ for small-amplitudes ($\delta \ll 1$) where $\theta=0,
\pi$ (the approximate $|\dot{E}_{\theta}|$ resembles the
approximate $|f_{x_{\theta}}|$ [1]). Only the geometry of the wave
occurs in the approximate $f$ and $\dot{E}$, however, we should
recall that (in page 2 of [1]) "In a free-swimming situation, the
upper sheet would move at a rate such that {\it the viscous drag
would balance with $f_x$}...". It means the statement "the
geometry of the wave is the only factor determining the direction
in which the relative position of the sheets evolve" is not
complete or valid! The energy dissipation or viscous dissipation
also plays an important role (otherwise the shape of the waveform
($a$) will be difficult to be preserved during the motions)! Note
that in [5] "... the energy of the beat is not only used for
propulsion, but also to overcome the sliding friction." is
claimed. Similar claims could also be traced in [6] although it
was not stated explicitly.

\end{document}